\documentclass[conference]{IEEEtran}
\IEEEoverridecommandlockouts
\usepackage[T1]{fontenc}
\usepackage[utf8]{inputenc}
\usepackage{cite}
\usepackage{amsmath,amssymb,amsfonts}
\usepackage{graphicx}
\usepackage{textcomp}
\usepackage{booktabs,array}
\usepackage{placeins}
\usepackage[hyphens]{url}
\usepackage{microtype}
\usepackage[hidelinks]{hyperref}
\hypersetup{pdftitle={Advisor, Co-pilot, or Guardian? AI Role Framing and Human Decision Behavior in Automated Vehicle Takeovers},pdfauthor={Wei-Hsiang Lo, Jingyi Xie, Gaojian Huang},pdfsubject={Accepted for publication in Journal of Safety Research},pdfkeywords={Human-AI Collaboration, Automated Driving, Control Authority, Explanation Content, Directional Semantics}}
\begin{document}
\title{Advisor, Co-pilot, or Guardian? AI Role Framing and Human Decision Behavior in Automated Vehicle Takeovers%
\thanks{Corresponding author: Gaojian Huang (gaojian.huang@sjsu.edu). Wei-Hsiang Lo is now with the University of Michigan.}}
\author{\IEEEauthorblockN{Wei-Hsiang Lo, Jingyi Xie, and Gaojian Huang}
\IEEEauthorblockA{Department of Industrial and Systems Engineering\\
San Jose State University, San Jose, CA, USA}}
\maketitle
\thispagestyle{plain}
\begin{abstract}
Artificial intelligence (AI) technology has gradually been integrated into automated vehicles. However, how this technology could systematically support takeover transitions in SAE Level 3 automated vehicles remains unknown, even though this level may remain a major form of driving automation in the near future. To address this, we implemented the three AI roles proposed by the triadic Human-AI collaboration framework (Advisor, Co-pilot, and Guardian) as potential authority-ability configurations for AI agents within an automated system. We then examined their interactions with explanation type (why and action) and directional semantics (ipsilateral and contralateral) in an automated driving simulator study with 24 participants. Results showed that, across the three AI roles, Advisor had the shortest decision-making time, followed by Guardian and Co-pilot. Advisor and Guardian were associated with higher decision-making accuracy and higher satisfaction ratings than Co-pilot. Guardian was associated with the lowest workload, followed by Advisor and Co-pilot. The effect of explanation type on decision-making time depended on AI role. Action explanations were faster under Advisor, whereas why explanations were faster under Co-pilot. For directional semantics, ipsilateral semantics were associated with shorter decision-making time than contralateral semantics. Overall, these findings provide preliminary guidance for designing AI-role authority-ability configurations and their associated speech-based takeover messages.
\end{abstract}
\begin{IEEEkeywords}
Human-AI Collaboration, Automated Driving, Control Authority, Explanation Content, Directional Semantics
\end{IEEEkeywords}
\vspace{0.6em}
\noindent\fbox{\begin{minipage}{\dimexpr\linewidth-2\fboxsep-2\fboxrule\relax}
\footnotesize
\textbf{Accepted manuscript.} Accepted for publication in \emph{Journal of Safety Research} on October 27, 2026. This is the author-prepared manuscript; the journal version is forthcoming.
\par\smallskip
\copyright\ 2026. This manuscript version is made available under the CC BY-NC-ND 4.0 license: \url{https://creativecommons.org/licenses/by-nc-nd/4.0/}.
\end{minipage}}
\vspace{0.6em}
\section{Introduction}\label{sec:1}

\subsection{Human-AI interaction in automated driving}\label{sec:1.1}

Artificial intelligence (AI) technology is rapidly evolving, as reflected in recent industry deployments. For example, Tesla has integrated Grok, a generative AI chatbot, into its vehicles \cite{r148}. Volkswagen \cite{r257}, Peugeot \cite{r222}, and Citro\"{e}n \cite{r147} plan to integrate ChatGPT into their future vehicle models. A potential reason for this adoption is that these technologies can enable systems to use data-driven prediction \cite{r136} and natural-language interaction \cite{r174} to interpret complex traffic scenes and provide real-time guidance to human drivers \cite{r244}. These capabilities can benefit current vehicle systems and may also provide a promising solution to the safety challenges at SAE Level 3, which may remain a major level of driving automation for a long time because of technical and ethical issues \cite{r135,r242}. At this level, the automated driving system is able to perform the entire dynamic driving task (DDT) within its operational design domain, but it still requires the human driver to intervene when the system reaches its operational limit. This intervention is known as a takeover, a cognitively and physically demanding process that gives the driver only a few seconds to correctly analyze the environment and perform an appropriate action. With capabilities to identify relevant hazards and anticipate how the situation may develop, AI agents operating during this transition are expected to provide timely support during the takeover response.

\subsection{AI roles in human-AI collaboration}\label{sec:1.2}

Before specifying how an AI agent may support a takeover transition at SAE Level 3, a conceptual distinction first needs to be made between technical AI ability and control authority. Technical ability refers to the tasks that an AI agent is capable of performing, and control authority refers to the level of control that an agent is allowed to exercise. Flemisch et al. \cite{r167} and Cavalcante Siebert et al. \cite{r141} both argued that it is important to align an agent's control authority with its technical ability. This principle means that the operational role of an AI agent within the automation system depends on how the abilities of agents are connected to the functions assigned to automation and how those abilities are exercised under the allocated control authority \cite{r219,r167}. For example, an AI agent may be technically capable of detecting a collision risk and initiating emergency braking. However, if the automation is authorized only to provide information, the agent may only warn the driver but cannot activate the brakes. In the context of SAE Level 3 automated vehicles, although SAE J3016 defines the division of responsibility during a takeover transition \cite{r228}, it is a taxonomy of driving automation and does not provide a framework for defining control authority at the level of individual takeover-support functions. Specifically, it does not specify how much control a takeover-support system is permitted to exercise, such as whether it may provide only a warning, assist with braking while the driver retains control, or initiate braking without driver confirmation.

Previous human-automation research has helped address this gap by defining different dimensions of function and authority allocation. The Types and Levels of Automation framework specifies which information, decision, and action functions are assigned to automation and to what degree \cite{r234,r219}. Adaptive automation specifies when this allocation may change \cite{r185}. Shared- and traded-control research specifies whether the human and automation exercise control concurrently or alternately \cite{r167,r152}. With these frameworks helping to define the function-level boundaries within which automation may exercise authority, Huang et al. \cite{r180} developed the triadic Human-AI collaboration framework to define the operational roles that AI can assume. Table~\ref{tab:1} compares the primary design dimension defined by each framework and the relationship between these boundaries and the operational roles.

\begin{table*}[!t]
\centering
\caption{Relationship between established automation frameworks and the triadic AI-role framework.}
\label{tab:1}
\begingroup
\footnotesize
\setlength{\tabcolsep}{3pt}
\renewcommand{\arraystretch}{1.15}
\begin{tabular}{@{}>{\raggedright\arraybackslash}p{0.23\linewidth}>{\raggedright\arraybackslash}p{0.27\linewidth}>{\raggedright\arraybackslash}p{0.46\linewidth}@{}}
\toprule
\textbf{Concept} & \textbf{Primary element defined} & \textbf{Relation to the triadic framework} \\
\midrule
SAE driving-automation levels \cite{r228} & Expected roles in sustained DDT performance and DDT fallback & Defines the system-level driving-automation boundary but not the real-time collaboration role of AI. \\
\addlinespace[3pt]
Types and Levels of Automation (\cite{r219}; extending \cite{r234}) & Automation type or function and the level of automation applied within each function & Defines what functions are automated and to what extent. \\
\addlinespace[3pt]
Adaptive automation \cite{r185} & Changes in function or control allocation over time & Defines when the allocation may change. \\
\addlinespace[3pt]
Shared and traded control \cite{r167,r152} & Concurrent or alternating distribution of control between the human and automation & Defines how control is distributed during task execution. \\
\addlinespace[3pt]
Ability-authority-responsibility alignment \cite{r167,r141} & Consistency between an agent's ability, authority, and responsibility & Provides the constraint that assigned authority should not exceed ability. \\
\addlinespace[3pt]
Triadic AI-role framework \cite{r180} & The real-time operational role of AI relative to the human & Connects AI abilities to functions and control-authority boundaries defined by established automation frameworks, yielding the Advisor, Co-pilot, and Guardian operational roles. \\
\bottomrule
\end{tabular}
\endgroup
\end{table*}

\begin{figure*}[!t]
\centering
\includegraphics[width=\linewidth]{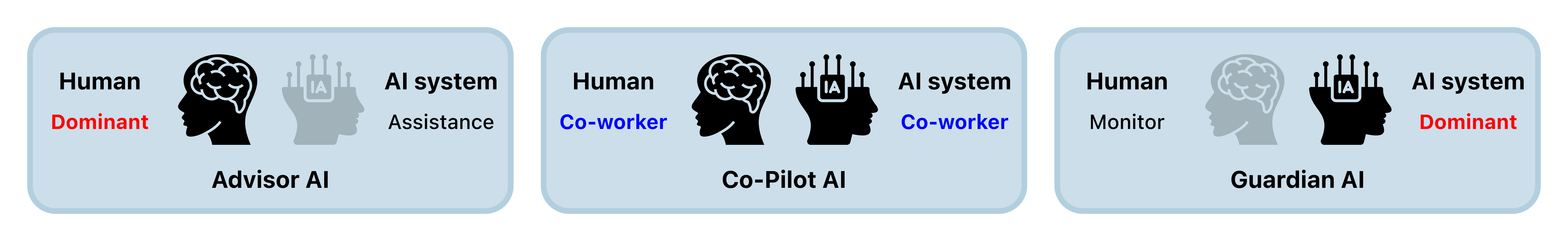}
\caption{Human-AI collaboration framework developed by Huang et al. \cite{r180}.}
\label{fig:1}
\end{figure*}

The three non-hierarchical operational roles proposed by Huang et al. \cite{r180} differ in the automation functions they support and the control authority under which AI abilities are exercised (Fig.~\ref{fig:1}). For Advisor AI, control authority is limited to the information level, including acquisition, analysis, and communication. Primary decision and action authority remain with the human driver. Under this authority arrangement, Advisor AI uses environmental perception and driver-state estimation to identify relevant hazards and traffic conditions and provide critical traffic information and guidance to help the driver make an appropriate decision during the takeover. For Co-pilot AI, control authority covers shared action implementation, allowing the driver and AI to contribute concurrently to vehicle control. Within this authority, Co-pilot uses environmental perception and prediction to generate steering, braking, or speed-control assistance when the driver's input is insufficient. For Guardian AI, control authority temporarily extends to full action implementation when a time-critical safety event occurs. Within this authority, the human driver still processes the message and indicates the selected lane but does not manually execute the maneuver. Guardian uses data-based risk assessment and maneuver-generation abilities to perform the emergency response, such as a lane change or braking, at that moment. Therefore, the three proposed roles represent three AI-role configurations that pair specific AI abilities with different degrees of control authority across automation functions.

Existing vehicle functions that incorporate partial forms of these three AI-role configurations have demonstrated safety benefits. Forward collision warning, which resembles the information support provided by Advisor, reduces rear-end striking crashes by 27\% \cite{r146}. Lane-keeping assistance, a function associated with Co-pilot, is estimated to reduce the risk of fatal single-vehicle road-departure crashes by 24\% \cite{r230}. Automatic emergency braking, a function associated with Guardian, reduces rear-end striking crash involvement by approximately 50\% \cite{r146}. This evidence supports the practical value of further developing these AI-role configurations with more advanced AI abilities. However, the three AI roles have not yet been empirically compared as complete authority-ability configurations within the same SAE Level 3 takeover context based on takeover performance, which reflects drivers' behavioral responses to AI support, and subjective ratings, which may be associated with their willingness to use these AI agents.

\subsection{Why-action information processing}\label{sec:1.3}

As the three AI-role configurations involve three different forms of cooperation between the system and the driver in action execution, the support provided under each configuration may need to be explained differently to the driver. For example, the communication may need to clarify whether the driver is responsible for selecting and performing the response under Advisor, performing it with assistance from Co-pilot, or selecting the response while Guardian performs the maneuver. Earlier studies have shown that such communication can reduce uncertainty and support timely driver responses \cite{r140,r159,r193}. However, when the reasoning process or limitations of AI are conveyed unclearly to the driver (e.g., waiting too long to take over or disengaging too early), the driver might place inappropriate trust in it \cite{r190,r196}. AI-related features often appear as a ``black box'' \cite{r157,r171}, which can make it difficult for the driver to understand what the system is doing, why it wants to do it, and what action it will take next. Previous studies show that when users understand how the system works, including its perception and intention, they can build an accurate mental model and calibrate their trust \cite{r134,r133,r232}. When a system provides explanations that reduce uncertainty and increase transparency between the individual and the system, users are more willing to adopt it \cite{r161,r212}. For example, Du et al. \cite{r158} presented participants with four types of explanations: (1) NExpl, in which the system did not explain its actions; (2) BExpl, in which explanations were provided 7 seconds before the automated vehicle's actions; (3) AExpl, in which explanations were provided within 1 second after the automated vehicle had acted; and (4) PermReq, in which the automated vehicle explained its upcoming action and asked the driver to approve or disapprove the action. Their results showed that providing an explanation, especially before action (BExpl), led to higher trust and preference than no explanation (NExpl). Additionally, when receiving information in an automated driving environment, drivers might prefer different communication styles, and this preference might lead to better performance. For example, Forster et al. \cite{r168} compared two types of takeover requests, one with nonspecific warning tones (i.e., generic), and the other with additional semantic speech output (e.g., speech + generic). Both subjective ratings (e.g., System Usability Scale) and objective performance (e.g., information processing time) indicated that the takeover request with additional semantic speech performed better than the request that used only nonspecific warning tones. That is, the effectiveness of human-AI communication depends on clearly presenting the intention of AI to increase transparency between humans and systems and on presenting information in a form that aligns with driver preferences and supports rapid and accurate interpretation under time pressure. Therefore, understanding whether the explanation aligns with human information processing is important.

There are two main cognitive routes that individuals use to process information \cite{r153,r186,r238}: (1) top-down processing, where individuals process information using their prior knowledge, goals, and expectations, and (2) bottom-up processing, where individuals process information through immediate sensory cues. Earlier findings suggest that an effective explanation design should align with these processing routes \cite{r138,r198,r213}. In takeover contexts, the system can provide a ``why'' explanation, which explains the reason behind a particular system behavior and enables the driver to interpret it based on their prior knowledge or experience, and thus triggers top-down processing. Conversely, to trigger bottom-up processing, the system can offer an ``action'' explanation, providing information about the system's intention to help individuals interpret the information based on immediate cues they collect, and thus facilitating the driver's information collection through those cues. However, the effectiveness of explanations in previous studies has been mixed. For example, Koo et al. \cite{r187} compared auditory takeover requests containing a combination of ``why'' (``Obstacle ahead.'') and ``action'' (``Car is braking.'') explanation messages. The results showed that when the takeover request included a ``why'' message, machine trust and acceptance were higher, whereas when the request included only an ``action'' (``how'') message, driver performance was worse (e.g., more road-edge excursions). Du et al. \cite{r159} also compared similar types of explanation messages but reported different results. Their comparison showed that the takeover request with a ``what will'' (action) message (e.g., ``Change to right lane'') was associated with a higher ease-of-use score than the ``why''-only message (e.g., ``Swerving vehicle ahead''). The conflicting findings may have arisen partly because the two studies evaluated explanation content under different allocations of task authority. In Koo et al. \cite{r187}, participants were required to continue driving, and the system applied emergency braking when a critical event occurred. Therefore, the control authority of the system was limited to intervention during critical events. In contrast, in Du et al. \cite{r159}, the system was responsible for most of the driving task, and drivers resumed control only when needed. In this case, the system remained engaged most of the time and transferred control to the driver when it reached its operational limit. The different automation functions permitted under these control-authority configurations may help explain the differing effects of explanation content. Such differences in explanation effects may also emerge across the three authority-ability configurations used in this study. In Advisor, drivers retain both decision and action authority; in Co-pilot, they share action implementation with the AI; and in Guardian, they identify the lane before the AI executes the immediate maneuver. Therefore, it is important to examine whether the effects of the two types of explanation content vary across these configurations within the same simulator context.

\subsection{Directional semantics}\label{sec:1.4}

Human-machine interfaces (HMIs) are important components of automated vehicles because they communicate system support to the driver, such as the support provided by each AI-role configuration implemented in the current study. Among different types of HMIs (e.g., head-up visual displays, speech-based auditory messages, and seat-based vibrotactile signals), speech-based auditory messages that deliver glance-free warnings could reduce visual and manual demands and remain effective even when drivers' posture changes \cite{r210,r226}. This may make speech a better choice than other modalities. For example, a visual display, such as a head-up display, often requires the driver to shift gaze to a specific location and visually search for the alert \cite{r201,r245}. This process might slow the driver's reaction. On the other hand, a vibrotactile display, such as a vibration cue sent through the seat back or steering wheel, may be less reliable when contact is inconsistent, such as when the driver's hands are not firmly on the steering wheel or the driver is wearing thick clothing \cite{r203,r211}. Such inconsistent contact can cause the driver to miss the signal. The widespread adoption of speech-based auditory messages in production vehicles \cite{r142,r178,r210} could also reduce bias arising from differences in individuals' prior knowledge of HMIs. Therefore, speech-based auditory messages might be a suitable modality for presenting the three AI-role configurations in the context of takeover support.  As the three AI-role configurations represent different allocations of action authority, directional information in speech-based takeover requests is expected to affect how drivers interpret and respond to the support provided by each configuration. In a lane-change takeover context, the directional reference may affect how drivers map the message onto a maneuver that they execute under Advisor, coordinate with AI assistance under Co-pilot, or select before the AI executes it under Guardian. These effects may depend on cognitive mechanisms for processing spatial information.  Based on stimulus-response compatibility (SRC) \cite{r165,r223}, two types of directional semantics have been explored for takeover requests in automated-vehicle takeover contexts: (1) ipsilateral semantics, in which the information offered aligns with the corresponding direction \cite{r203,r211,r220}, and (2) contralateral semantics, in which the information provided is contrary to the compatibility direction \cite{r145,r182,r208}. In a takeover scenario where a sudden obstacle appears, ipsilateral semantics would indicate a direction to steer away from the hazard (e.g., corresponding with the same direction as the safer lane). Meanwhile, contralateral semantics would inform the driver of the obstacle's location.

Previous studies indicated inconsistent effects of these two directional semantics. Arabian et al. \cite{r128} compared toward-free-lane and toward-hazard spoken warnings in an SAE Level 2 driving simulator. Participants received the words ``Left'' and ``Right'' through the corresponding ear with different onset intervals between the auditory warning and visual information. The results indicated that toward-free-lane warnings produced faster steering responses than toward-hazard warnings. However, de Winter et al. \cite{r151} evaluated these two directional semantics, namely ipsilateral spoken warnings (``Go Left/Right'') and contralateral spoken warnings (``Danger Left/Right''), using animated near-collision videos and asked participants to make their lane-change decisions using the left or right arrow keys. Although participants rated ipsilateral warnings as more satisfactory, the objective results revealed no significant differences in either response time or accuracy. Two differences in the study designs are relevant. Arabian et al. \cite{r128} embedded the warnings in a hands-on SAE Level 2 simulator drive under time pressure, whereas de Winter et al. \cite{r151} presented video clips to seated participants without vehicle control. The latter setting did not include the urgency or motor demands of a takeover. The response measures also differed. Arabian et al. used steering input, while de Winter et al. used a key press. In addition, both studies presented the directional word from the corresponding side, so spoken meaning and sound location varied together. As a result, their separate contributions cannot be determined. The current study addressed these limitations by testing directional wording within an SAE Level 3 takeover under time pressure. The sound source was held constant, and the same indicator-button response was used in every condition. This design isolated the effect of spoken semantics within each allocation of action authority.

\subsection{Current study}\label{sec:1.5}

To address these gaps, the current study aimed to empirically compare the three AI roles as complete authority-ability configurations (Advisor, Co-pilot, and Guardian) within the same SAE Level 3 takeover context and investigate the effects of explanation type (why and action) and directional semantics (ipsilateral and contralateral) on objective takeover performance and subjective ratings (i.e., trust, anxiety, usefulness, satisfaction, and workload). The study addressed the following research questions:

RQ1. How do the three AI-role authority-ability configurations differ in objective takeover performance and subjective ratings?

RQ2. How does explanation type affect objective takeover performance and subjective ratings, and do these effects vary across the three AI-role configurations?

RQ3. How do spoken directional semantics affect objective takeover performance and subjective ratings, and do these effects vary across the three AI-role configurations?

Our study makes three main contributions. First, we empirically compared three AI-role configurations that combine technical AI abilities with different degrees of control authority in terms of objective performance and subjective ratings in automated driving contexts. Second, we outlined guidelines for activating appropriate cognitive mechanisms during time- and safety-critical scenarios across three AI-role configurations involving different forms of cooperation. Third, we investigated directional semantics conveyed through speech-based auditory messages across the three AI-role configurations. The findings of this study provide guidance on integrating AI-driven systems and enhancing transparency for human operators in next-generation intelligent systems.

\section{Method}\label{sec:2}

\subsection{Participants}\label{sec:2.1}

Twenty-four volunteers participated in this study. They were aged between 21 and 40 years (mean age = 25.38, standard deviation (SD) = 3.85). Participants reported driving an average of three days (SD = 1.69) per week in the past year. All participants were recruited from the research pool system at San Jose State University. To compensate for their time and participation, each participant received three hours of course credit. To participate in this study, volunteers were required to (1) possess a valid driver's license, (2) drive at least once a week, (3) have no cognitive or neurological impairments affecting hearing, and (4) report no motion sickness. The study received approval from the San Jose State University Institutional Review Board (IRB protocol \#: 21278). All participants provided informed consent prior to participation, and data were anonymized before analysis.

\subsection{Apparatus}\label{sec:2.2}

\begin{figure}[!t]
\centering
\includegraphics[width=\linewidth]{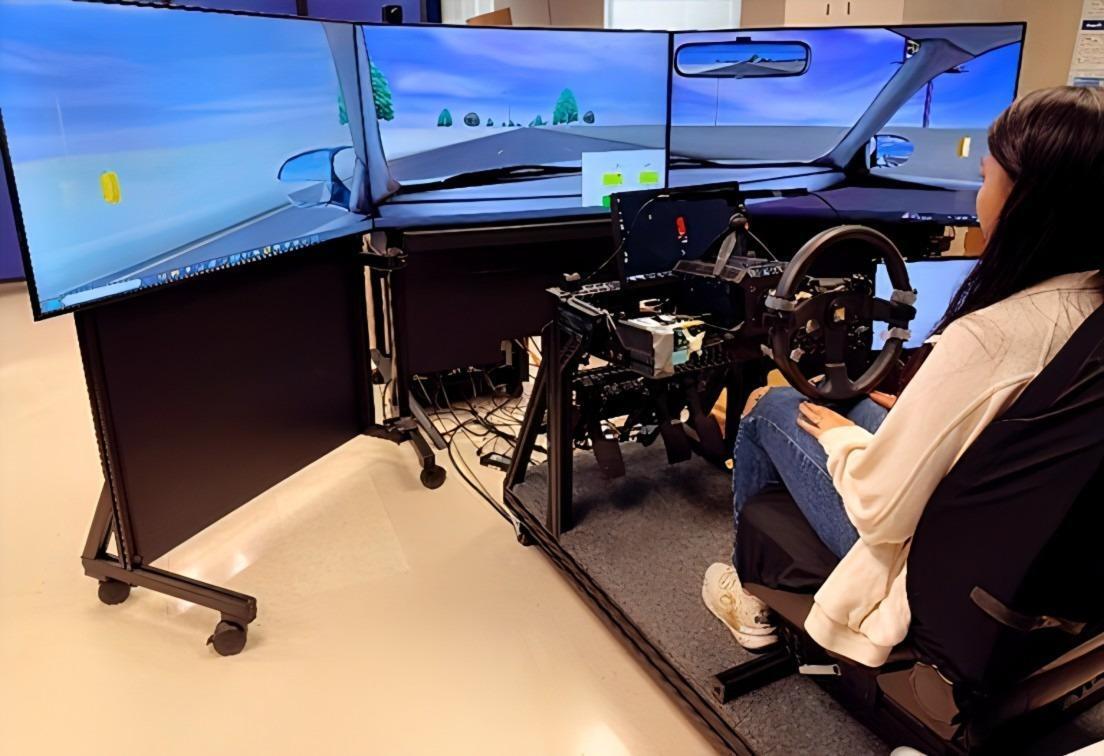}
\caption{The fixed-base medium-fidelity driving simulator, miniSim.}
\label{fig:2}
\end{figure}

The fixed-base medium-fidelity driving simulator, miniSim, was used in this study. This simulator was developed by the University of Iowa Driving Safety Research Institute (DSRI) (see Fig.~\ref{fig:2}). To simulate the driving environment, three 42-inch monitors were used for display, and an 18.5-inch monitor served as the in-vehicle dashboard. In addition, the simulator included accelerator and brake pedals, a steering wheel, and a standard-size adjustable driver's seat. The driving data were collected at a rate of 60 Hz.

\subsection{Speech-based takeover requests}\label{sec:2.3}

Speech-based takeover requests were delivered through a synthesized female voice, which was generated by Amazon Polly (name: Joanna, gender: female, nationality: USA). The synthesized voice has been widely used in previous studies \cite{r131,r132,r250}, and female voices have been associated with better takeover performance than male voices \cite{r130,r252}. To verify that participants could hear the speech-based takeover requests at an appropriate volume, we conducted a pilot study before the main experiment. Following testing, all speech-based takeover requests were amplified by 20 dB using Adobe Premiere Pro. Since the takeover process occurred under critical conditions, each speech-based message was limited to approximately two seconds \cite{r139,r251}, and for clarity, it was repeated once during each takeover event. To minimize potential bias and support effectiveness in time-critical scenarios, all speech-based takeover requests were standardized at a speech rate of 170 words per minute (wpm) \cite{r256}. Additionally, to reduce serial-position biases, including the primacy and recency effects (the former involves information presented earlier that tends to be more memorable, while the latter refers to information presented at the end that is easier to recall) \cite{r214,r233}, directional keywords were placed in the middle of speech-based takeover requests (see Table~\ref{tab:2}) to avoid a specific memorized pattern and mitigate expectancy-based responses. To isolate the effects of wording from spatial sound location, all messages were delivered through the same in-vehicle speaker at the same location.

\begin{table*}[!t]
\centering
\caption{The integration of AI roles, along with explanations and directional semantics for speech-based takeover requests.}
\label{tab:2}
\begingroup
\footnotesize
\setlength{\tabcolsep}{3pt}
\renewcommand{\arraystretch}{1.15}
\begin{tabular}{@{}>{\raggedright\arraybackslash}p{0.11\linewidth}>{\raggedright\arraybackslash}p{0.2075\linewidth}>{\raggedright\arraybackslash}p{0.2075\linewidth}>{\raggedright\arraybackslash}p{0.2075\linewidth}>{\raggedright\arraybackslash}p{0.2075\linewidth}@{}}
\toprule
\textbf{AI Role} & \textbf{Why + ipsilateral} & \textbf{Why + contralateral} & \textbf{Action + ipsilateral} & \textbf{Action + contralateral} \\
\midrule
Advisor & Hazard ahead, left/right lane clear. & Hazard ahead, left/right lane block. & Change to the left/right lane immediately. & Avoid the left/right immediately. \\
\addlinespace[3pt]
Co-pilot & Realign, left/right steering insufficient. & Realign, left/right side drift detected. & Hold steady, applying left/right steering. & Hold steady, reducing left/right steering. \\
\addlinespace[3pt]
Guardian & Step in, detected left/right lane clear. & Step in, detected left/right side risk. & Engaging, steering left/right for safety. & Engaging, avoiding left/right risk for safe. \\
\bottomrule
\end{tabular}
\endgroup
\end{table*}

\subsection{Driving scenario}\label{sec:2.4}

\begin{figure*}[!t]
\centering
\includegraphics[width=0.72\linewidth]{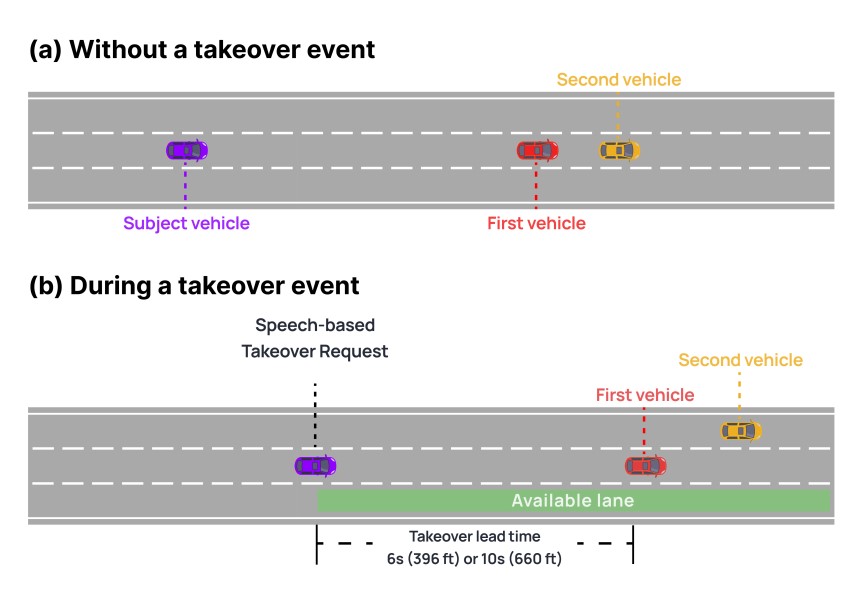}
\caption{An example bird's-eye view of the driving scenario \cite{r181,r203}: (a) Without a takeover event: the subject vehicle (purple) was driving autonomously on a three-lane highway with two vehicles traveling in front of the subject vehicle. (b) During a takeover event, the subject vehicle (purple) was expected to perform a right lane change to avoid collisions with the first vehicle (red), which stopped approximately 396 feet or 660 feet ahead, and with a second vehicle (yellow) that was changing lanes to the left.}
\label{fig:3}
\end{figure*}

During the experiment, participants rode in a simulated SAE Level 3 automated vehicle. At this level, the vehicle automatically performed the driving task unless a takeover was needed. During the experiment, the subject vehicle traveled on a three-lane highway and maintained its position in the center lane at a speed of 45 mph. This speed reflects reduced speed conditions common on urban freeways (e.g., work zones) \cite{r173,r243,r253}. Additionally, two vehicles traveled in front of the subject vehicle at a distance of approximately 792 feet (see Fig.~\ref{fig:3}). Occasionally, the first vehicle suddenly stopped in the center lane, which caused the automated system of the subject vehicle to fail \cite{r156,r177,r258}. Due to the obstacle detection feature, the automated system triggered a takeover request through the in-vehicle speaker when a certain distance was reached. To reduce the effect of varying urgency levels, the current study followed previous research and used two lead times that represented different urgency levels (i.e., approximately 396 ft, corresponding to a 6-second takeover lead time, and approximately 660 ft, corresponding to a 10-second lead time) \cite{r229,r241}.

When the first vehicle suddenly stopped, the second vehicle (i.e., the vehicle in front of the first vehicle) moved either to the left or the right lane to block one specific lane, leaving only one lane available to participants. The purpose was to increase the complexity of the driving environment and require participants to make appropriate decisions in critical scenarios. Therefore, participants in the current scenario needed to avoid the suddenly stopped vehicle, determine the movement of the second vehicle, and select the available lane to move into. Previous takeover studies have shown that where drivers look, how quickly they reposition themselves, and when their initial control inputs occur during a takeover directly influence their decision-making process. Including this process in the simulated study could improve its realism \cite{r181,r221}.

\begin{figure*}[!t]
\centering
\includegraphics[width=\linewidth]{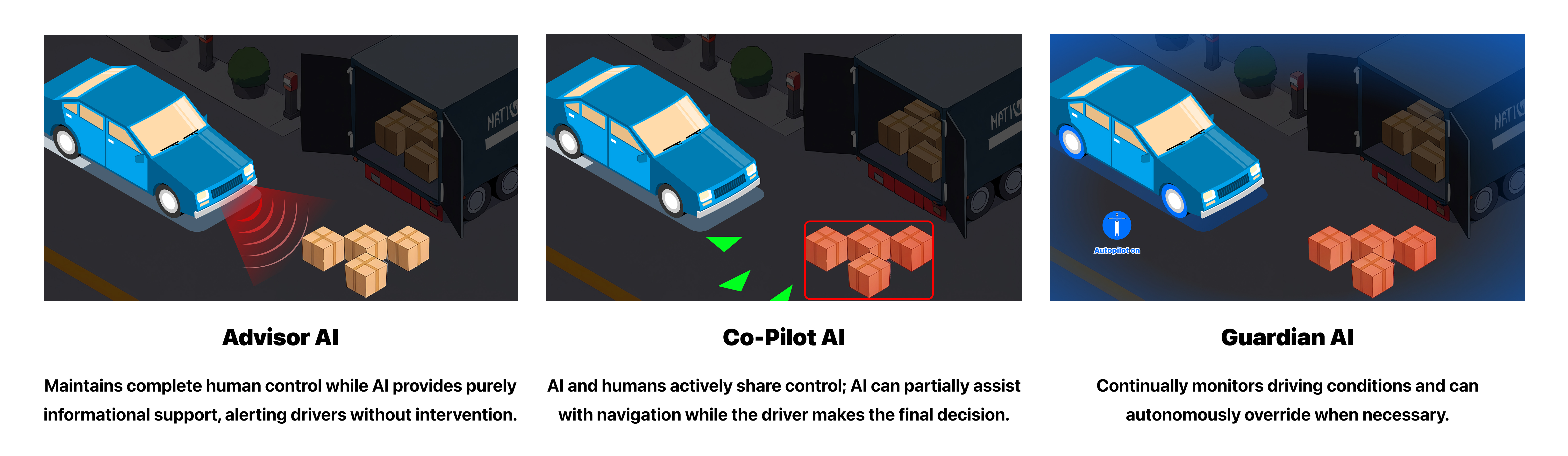}
\caption{Authority-ability configurations of the three AI roles.}
\label{fig:4}
\end{figure*}

After participants perceived the speech-based takeover request, they processed the message and traffic conditions. Because the three AI roles differed in action authority, participants performed different actions across configurations (Fig.~\ref{fig:4}; \cite{r180}). In every configuration, participants first selected the left or right lane and pressed the corresponding indicator button on the steering wheel \cite{r197,r209}. In the Advisor and Co-pilot configurations, they then pressed the brake to disengage automation before manually executing the maneuver. In the Guardian configuration, automation remained engaged, and the system subsequently executed the maneuver.

After completing the decision-making action, participants in the Advisor and Co-pilot configurations were required to manually drive in that lane and maintain a speed of 45 mph. Once they passed the obstacle (i.e., the stopped vehicle), they were instructed to return to the center lane and reactivate automation mode by pressing the button next to the steering wheel. Under the Guardian configuration, the system handled the lane change during the process, and no manual steering input from participants was required. After the vehicle passed the obstacles, it automatically returned to the center lane. Although the three AI-role configurations combined technical AI abilities with different degrees of control authority and thus required participants to perform different actions, the primary response measure was the same indicator-button response used to indicate lane choice in every condition.

\subsection{Procedure}\label{sec:2.5}

\begin{figure*}[!t]
\centering
\includegraphics[width=\linewidth]{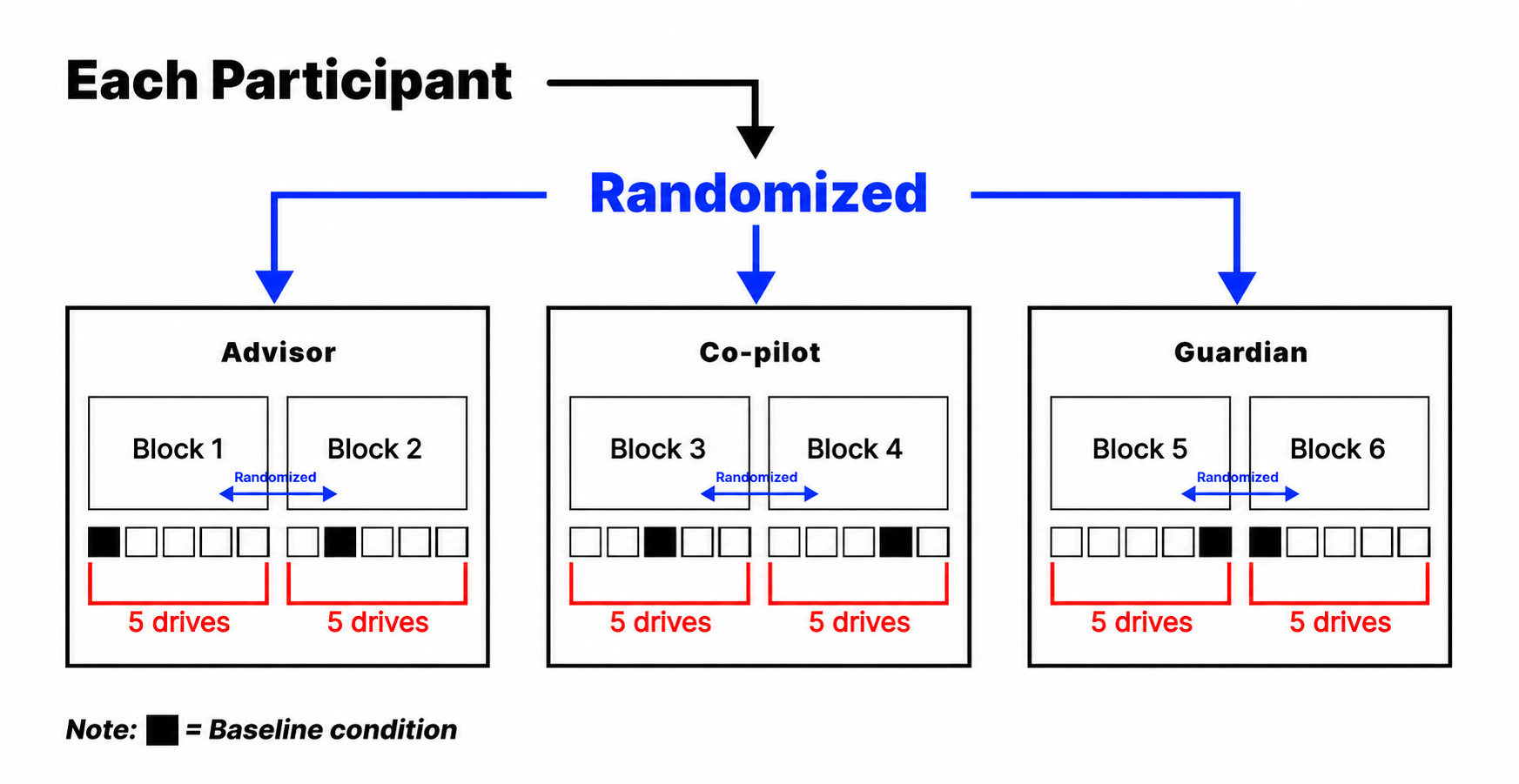}
\caption{An overview of the experiment structure (per participant).}
\label{fig:5}
\end{figure*}

Once participants arrived, they were asked to sign a consent form acknowledging their rights and providing consent to participate. Afterward, they completed the pre-experiment questionnaire, which collected both demographic information and their prior attitudes toward AI-related features (i.e., anxiety, trust, and preferences). Following these initial preparations, the experiment proceeded through the steps described below. Before the driving simulator session, participants completed a 20-minute training session. During training, they familiarized themselves with the simulator, the speech-based takeover requests, and the actions required during a takeover. In this within-subjects design, each participant completed drives under all three AI-role configurations after training (Advisor, Co-pilot, and Guardian; see Fig.~\ref{fig:5}). Each AI-role configuration contained two five-drive blocks, each comprising one Baseline drive and four experimental drives. Thus, participants completed 10 drives per AI-role configuration and 30 drives in total. To distribute potential order effects, the orders of the three AI-role configurations, the two blocks within each configuration, and the five drives within each block were randomized and balanced across participants (see Fig.~\ref{fig:5}). The baseline drive used the original equipment manufacturer (OEM)-style alert (i.e., the standard Tesla takeover warning sound). The four experimental drives represented the four combinations of explanation type (why or action) and directional semantics (ipsilateral or contralateral) shown in Table~\ref{tab:2}. In every drive, the average time from the start of the drive to the delivery of the takeover request was about one and a half minutes. Before each AI-role configuration began, the speech-based takeover requests for that configuration were presented to participants to confirm their understanding of the language and the intended meaning.

\begin{figure*}[!t]
\centering
\includegraphics[width=\linewidth]{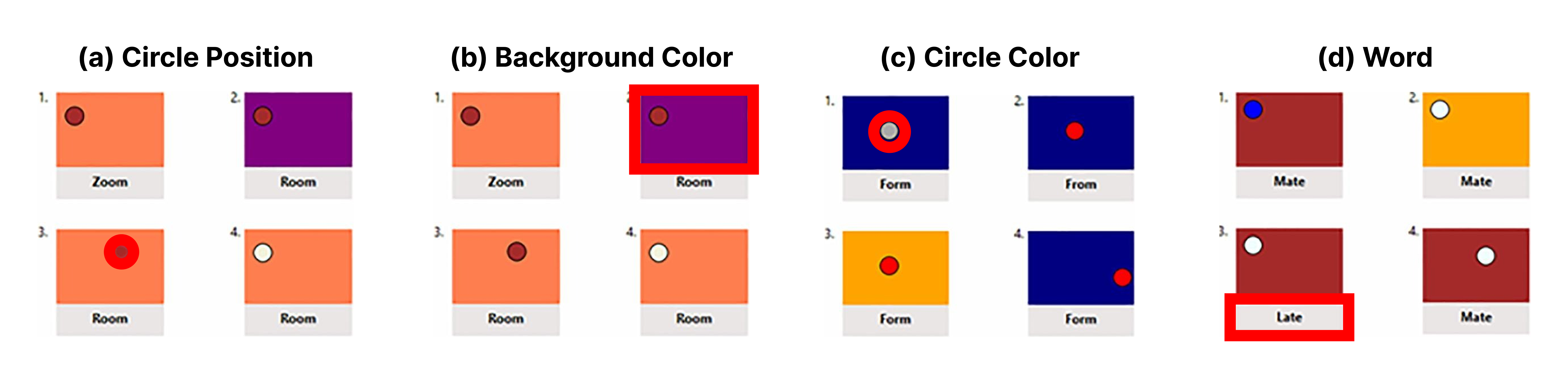}
\caption{Example of all four types of the ``Spot the Difference'' game: (a) circle position, (b) background color, (c) circle color, and (d) the spelling of words.}
\label{fig:6}
\end{figure*}

After each drive, participants were asked to report their subjective ratings of the implemented AI-role configuration and the speech-based takeover request they had just experienced, including trust, anxiety, usefulness, satisfaction, and perceived workload. There were 3-minute breaks between drives and 10-minute breaks between blocks to mitigate fatigue. Additionally, to reduce any early preparation for takeover, participants engaged in the ``Spot the Difference'' game (Fig.~\ref{fig:6}), which has been used to guide drivers' attention off the road \cite{r200,r246} and to simulate non-driving-related tasks that drivers might perform in real vehicles. The game window appeared in a corner of the display, showing four boxes with text prompts. In each round, participants identified which box was different from the others based on the prompt (e.g., circle position in Fig.~\ref{fig:6}(a), background color in Fig.~\ref{fig:6}(b), circle color in Fig.~\ref{fig:6}(c), and word spelling in Fig.~\ref{fig:6}(d)) shown at the top of the game window. After making the selection, participants verbally reported the answer, and the experimenter recorded the response. A new round then began. At the end of the session, after a total of 30 drives were completed, participants filled out the post-experiment questionnaire, which asked about their overall preference for all the speech-based takeover requests and their overall experience during the experiment. The whole experiment took approximately 180 minutes.

\subsection{Dependent measures}\label{sec:2.6}

Objective performance measures and subjective ratings were collected in this study. Decision-making time and decision-making accuracy served as the objective performance measures in the current study. Decision-making time was measured as the interval between the start of the takeover request and the moment participants pressed the indicator button \cite{r197,r202}. This measure was used to assess how quickly participants perceived and processed different speech-based takeover requests. Decision-making accuracy \cite{r151,r205} was measured by the proportion of takeover events in which participants identified the traffic situation and pressed the appropriate indicator button (left/right) corresponding to the scenario they encountered. For each participant, accuracy was calculated as the number of correct button presses divided by the total number of takeover events.

Participants also evaluated the experimental conditions using subjective ratings of trust, anxiety, usefulness, satisfaction, and workload. Trust was measured using six items rated on a 7-point Likert scale developed by Du et al. \cite{r158} for driver-automated vehicle (AV) interaction and adapted from Muir \cite{r215}: (1) competence, (2) predictability, (3) dependability, (4) responsibility, (5) reliability, and (6) faith. Specifically, rating 1 indicated ``not at all,'' and rating 7 indicated ``extremely well.'' A higher average score indicated greater trust in the system. We used this scale to compare participants' trust across our independent variables. The anxiety questionnaire was primarily adapted from a published model developed by the CHIMe Lab at Stanford University, which has been used to assess driver attitudes toward the in-vehicle system \cite{r189,r217,r240}. These items assessed participants' anxiety in a critical scenario, which could influence the trust-performance link in takeover tasks. The average response across the following four items was used to represent anxiety in an automated-system context: (1) fearful, (2) afraid, (3) anxious, and (4) uneasy.

Usefulness and satisfaction were rated based on the Technology Acceptance Questionnaire developed by van der Laan et al. \cite{r248}. This questionnaire contained nine items. Five items related to usefulness were: (1) Ease of Use, (2) General Quality, (3) Superfluity, (4) Effectiveness, and (5) Alertness Impact. Additionally, four items related to satisfaction were: (1) Pleasantness, (2) Enjoyability, (3) Likability, and (4) Desirability. The scores for usefulness and satisfaction were calculated by averaging responses to their respective items. These two ratings helped clarify participants' overall preferences regarding the implemented AI-role configuration and the speech-based takeover requests they had just experienced.

Cronbach's alpha estimates whether items on a multi-item scale consistently measure the same construct; higher values indicate greater internal consistency. Previous studies reported alpha values of .94 for trust, .91 for anxiety \cite{r188}, .73 to .91 for usefulness (median = .82; \cite{r264}), and .81 to .94 for satisfaction (median = .86; \cite{r264}).

Workload was measured with the NASA Task Load Index (NASA-TLX), a multidimensional self-report measure of perceived workload \cite{r176}. The NASA-TLX includes six subscales: (1) mental demand, (2) physical demand, (3) temporal demand, (4) performance, (5) effort, and (6) frustration. Participants were asked to rate each subscale from 0 to 10, and the NASA-TLX score was calculated by summing the scores across all subscales, resulting in a total score ranging from 0 to 60 \cite{r144}. Higher scores indicated greater perceived workload.

\subsection{Data analysis}\label{sec:2.7}

\subsubsection{Full-factorial analyses}\label{sec:2.7.1}

The study used a 3 (AI role: Advisor, Co-pilot, and Guardian) $\times$ 2 (explanation type: why and action) $\times$ 2 (directional semantics: ipsilateral and contralateral) full-factorial within-subjects design. We analyzed decision-making accuracy at the trial level using a generalized linear mixed model (GLMM) with a binomial distribution and logit link. The full-factorial model included AI role, explanation type, directional semantics, and all two- and three-way interactions as fixed effects, with a random intercept for participant. Because the factorial data exhibited separation and the ordinary maximum-likelihood model was nearly unidentifiable, we estimated the model using maximum penalized likelihood with weakly regularizing zero-centered normal priors (SD = 2.5) on the nonintercept fixed effects. For decision-making accuracy, we reported Type III Wald $\chi^2$ tests. For the comparison of the three AI roles with the sound-only Baseline, we additionally reported model-based predicted probabilities with 95\% confidence intervals (CIs) and Bonferroni-adjusted pairwise odds ratios (ORs). We analyzed decision-making time and the five subjective ratings using three-way repeated-measures analyses of variance (ANOVAs). Table~\ref{tab:3} maps each dependent variable to its full-factorial analysis and Baseline comparison.

To examine potential learning and fatigue effects, we conducted a temporal-order sensitivity analysis. First, we reconstructed the chronological order of the 30 drives from the participant-specific block sequence and within-block counterbalancing records. Second, we retained the 24 full-factorial drives at their actual positions within the 30-drive session and added standardized linear and quadratic trial-position terms to each factorial model. Continuous outcomes were analyzed using linear mixed-effects models with a random intercept for participant and a participant-specific linear trial slope, whereas decision-making accuracy was analyzed using a regularized binomial GLMM with a random intercept for participant. Holm corrections were applied across the seven outcomes.

\begin{table*}[!t]
\centering
\caption{Mapping of dependent variables to statistical analyses.}
\label{tab:3}
\begingroup
\footnotesize
\setlength{\tabcolsep}{3pt}
\renewcommand{\arraystretch}{1.15}
\begin{tabular}{@{}>{\raggedright\arraybackslash}p{0.19\linewidth}>{\raggedright\arraybackslash}p{0.40\linewidth}>{\raggedright\arraybackslash}p{0.35\linewidth}@{}}
\toprule
\textbf{Dependent variable} & \textbf{Full-factorial analysis} & \textbf{Baseline comparison} \\
\midrule
Decision-making time & Three-way repeated-measures ANOVA (AI role $\times$ explanation type $\times$ directional semantics) & One-way repeated-measures ANOVA (Advisor, Co-pilot, Guardian, and Baseline) \\
\addlinespace[3pt]
Decision-making accuracy & Binomial GLMM with a logit link (AI role $\times$ explanation type $\times$ directional semantics) & Binomial GLMM with a logit link (Advisor, Co-pilot, Guardian, and Baseline) \\
\addlinespace[3pt]
Trust & Three-way repeated-measures ANOVA (AI role $\times$ explanation type $\times$ directional semantics) & Not analyzed \\
\addlinespace[3pt]
Anxiety & Three-way repeated-measures ANOVA (AI role $\times$ explanation type $\times$ directional semantics) & Not analyzed \\
\addlinespace[3pt]
Usefulness & Three-way repeated-measures ANOVA (AI role $\times$ explanation type $\times$ directional semantics) & Not analyzed \\
\addlinespace[3pt]
Satisfaction & Three-way repeated-measures ANOVA (AI role $\times$ explanation type $\times$ directional semantics) & Not analyzed \\
\addlinespace[3pt]
Workload & Three-way repeated-measures ANOVA (AI role $\times$ explanation type $\times$ directional semantics) & Not analyzed \\
\bottomrule
\end{tabular}
\endgroup
\end{table*}

\subsubsection{Baseline comparisons}\label{sec:2.7.2}

We compared the three AI roles with the Baseline condition, which used an OEM-style alert (i.e., the Tesla takeover sound) without AI-related features (see Fig.~\ref{fig:5}). This comparison tested whether each AI role improved objective performance relative to current takeover practice. For decision-making time, we used a one-way repeated-measures ANOVA to compare the average time per drive across Advisor, Co-pilot, Guardian, and Baseline for each participant. We compared decision-making accuracy across the three AI roles and the Baseline using a GLMM with a binomial distribution, a logit link, and a random intercept for participant. As the Baseline condition served as the control for the authority-ability configurations and used a fixed tone that did not include any explanation or directional-semantics content, the current study treated it as an additional control in an augmented factorial design, in which main and interaction effects are estimated within the factorial treatment set and the additional control is evaluated through separate comparisons \cite{r265,r266}. Because of the experimental design, objective measures such as decision-making time and accuracy were based on direct participant behavior using the same indicator-button response in every drive. However, because the Baseline had no equivalent combination of AI role, explanation type, and directional semantics, participants might have used a different reference when providing subjective ratings, which could bias the comparison. We consequently limited the Baseline comparison to objective performance.

\subsubsection{Statistical corrections and decision criteria}\label{sec:2.7.3}

We used Greenhouse-Geisser estimates to correct the degrees of freedom when the data violated sphericity. We applied Bonferroni corrections to multiple comparisons across all levels. We set the significance threshold at $p$ \textless{} 0.05 and reported effect sizes using partial eta squared ($\eta_p^2$) for omnibus ANOVA effects, paired Cohen's $d_z$ for continuous pairwise comparisons, and ORs for accuracy comparisons. Continuous pairwise comparisons were also reported as mean differences (MDs) with 95\% CIs.

\section{Results}\label{sec:3}

Table~\ref{tab:4} summarizes all prespecified two- and three-way interaction tests for the continuous outcomes. Outcome-specific results are reported below.

\begin{table*}[!t]
\centering
\caption{Two- and three-way interaction tests for continuous outcomes.}
\label{tab:4}
\begingroup
\footnotesize
\setlength{\tabcolsep}{3pt}
\renewcommand{\arraystretch}{1.08}
\begin{tabular}{@{}>{\raggedright\arraybackslash}p{1.15in}>{\raggedright\arraybackslash}p{2.65in}rrrrr@{}}
\toprule
\textbf{Outcome} & \textbf{Interaction} & \textbf{$df_1$} & \textbf{$df_2$} & \textbf{F} & \textbf{p} & \textbf{$\eta_p^2$} \\
\midrule
Decision-making time & AI role $\times$ explanation type & 1.728 & 39.753 & 15.654 & \textless{} .001 & .405 \\
Decision-making time & AI role $\times$ directional semantics & 1.710 & 39.325 & .035 & .949 & .002 \\
Decision-making time & Explanation type $\times$ directional semantics & 1.000 & 23.000 & 2.346 & .139 & .093 \\
Decision-making time & AI role $\times$ explanation type $\times$ directional semantics & 1.832 & 42.125 & 1.910 & .164 & .077 \\
\addlinespace[3pt]
Trust & AI role $\times$ explanation type & 1.650 & 37.951 & .168 & .805 & .007 \\
Trust & AI role $\times$ directional semantics & 1.842 & 42.366 & .340 & .696 & .015 \\
Trust & Explanation type $\times$ directional semantics & 1.000 & 23.000 & .002 & .964 & .000 \\
Trust & AI role $\times$ explanation type $\times$ directional semantics & 1.730 & 39.795 & 1.064 & .346 & .044 \\
\addlinespace[3pt]
Anxiety & AI role $\times$ explanation type & 1.588 & 36.526 & .255 & .725 & .011 \\
Anxiety & AI role $\times$ directional semantics & 1.855 & 42.667 & 4.955 & .013 & .177 \\
Anxiety & Explanation type $\times$ directional semantics & 1.000 & 23.000 & 1.856 & .186 & .075 \\
Anxiety & AI role $\times$ explanation type $\times$ directional semantics & 1.608 & 36.989 & 1.311 & .277 & .054 \\
\addlinespace[3pt]
Usefulness & AI role $\times$ explanation type & 1.996 & 45.901 & .309 & .735 & .013 \\
Usefulness & AI role $\times$ directional semantics & 1.999 & 45.966 & .074 & .929 & .003 \\
Usefulness & Explanation type $\times$ directional semantics & 1.000 & 23.000 & .159 & .694 & .007 \\
Usefulness & AI role $\times$ explanation type $\times$ directional semantics & 1.915 & 44.039 & 1.390 & .260 & .057 \\
\addlinespace[3pt]
Satisfaction & AI role $\times$ explanation type & 1.714 & 39.426 & .175 & .807 & .008 \\
Satisfaction & AI role $\times$ directional semantics & 1.900 & 43.700 & .173 & .831 & .007 \\
Satisfaction & Explanation type $\times$ directional semantics & 1.000 & 23.000 & .505 & .484 & .021 \\
Satisfaction & AI role $\times$ explanation type $\times$ directional semantics & 1.917 & 44.101 & 1.166 & .319 & .048 \\
\addlinespace[3pt]
Workload & AI role $\times$ explanation type & 1.592 & 36.624 & 2.608 & .098 & .102 \\
Workload & AI role $\times$ directional semantics & 1.803 & 41.467 & .564 & .556 & .024 \\
Workload & Explanation type $\times$ directional semantics & 1.000 & 23.000 & .010 & .923 & .000 \\
Workload & AI role $\times$ explanation type $\times$ directional semantics & 1.837 & 42.260 & 1.403 & .257 & .057 \\
\bottomrule
\end{tabular}
\endgroup
\end{table*}

\subsection{Decision-making time}\label{sec:3.1}

\subsubsection{Full-factorial analysis}\label{sec:3.1.1}

\begin{figure*}[!t]
\centering
\includegraphics[width=\linewidth]{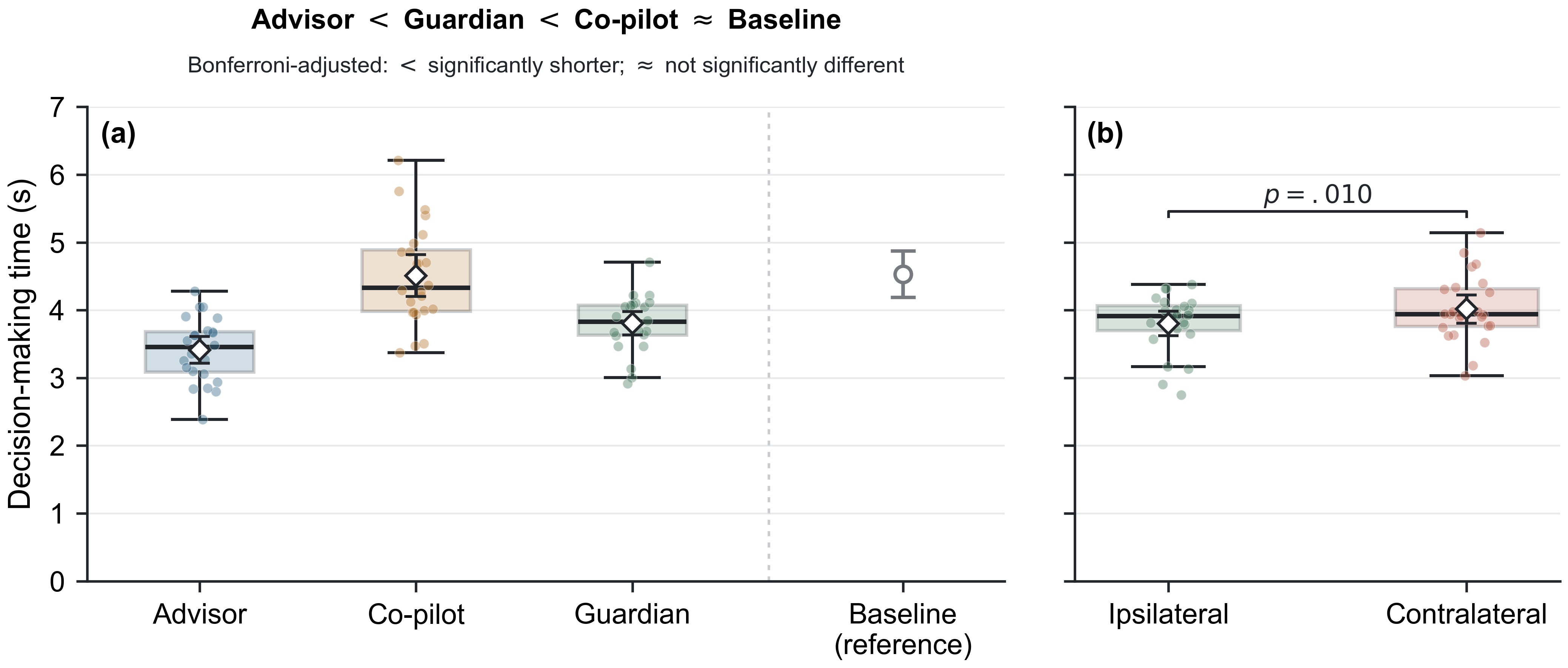}
\caption{Decision-making time across (a) the three AI roles and Baseline, and (b) the two directional semantics conditions.}
\label{fig:7}
\end{figure*}

There were significant main effects of AI role ($F$ (1.519, 34.942) = 39.178, $p$ \textless{} .001, $\eta_p^2$ = .630; Fig.~\ref{fig:7}(a)) and directional semantics ($F$ (1, 23) = 7.880, $p$ = .010, $\eta_p^2$ = .255) (Fig.~\ref{fig:7}(b)) on decision-making time. Across AI roles, Advisor (mean (M) = 3.413 s, standard error of the mean (SEM) = .095) had the shortest decision-making time, followed by Guardian ($M$ = 3.805 s, SEM = .084). Co-pilot ($M$ = 4.510 s, SEM = .149) had the longest decision-making time. For directional semantics, ipsilateral semantics ($M$ = 3.803 s, SEM = .088) were associated with shorter decision-making time than contralateral semantics ($M$ = 4.016 s, SEM = .101). There was no significant main effect of explanation type ($F$ (1, 23) = .860, $p$ = .363, $\eta_p^2$ = .036) on decision-making time.

\begin{figure*}[!t]
\centering
\includegraphics[width=0.82\linewidth]{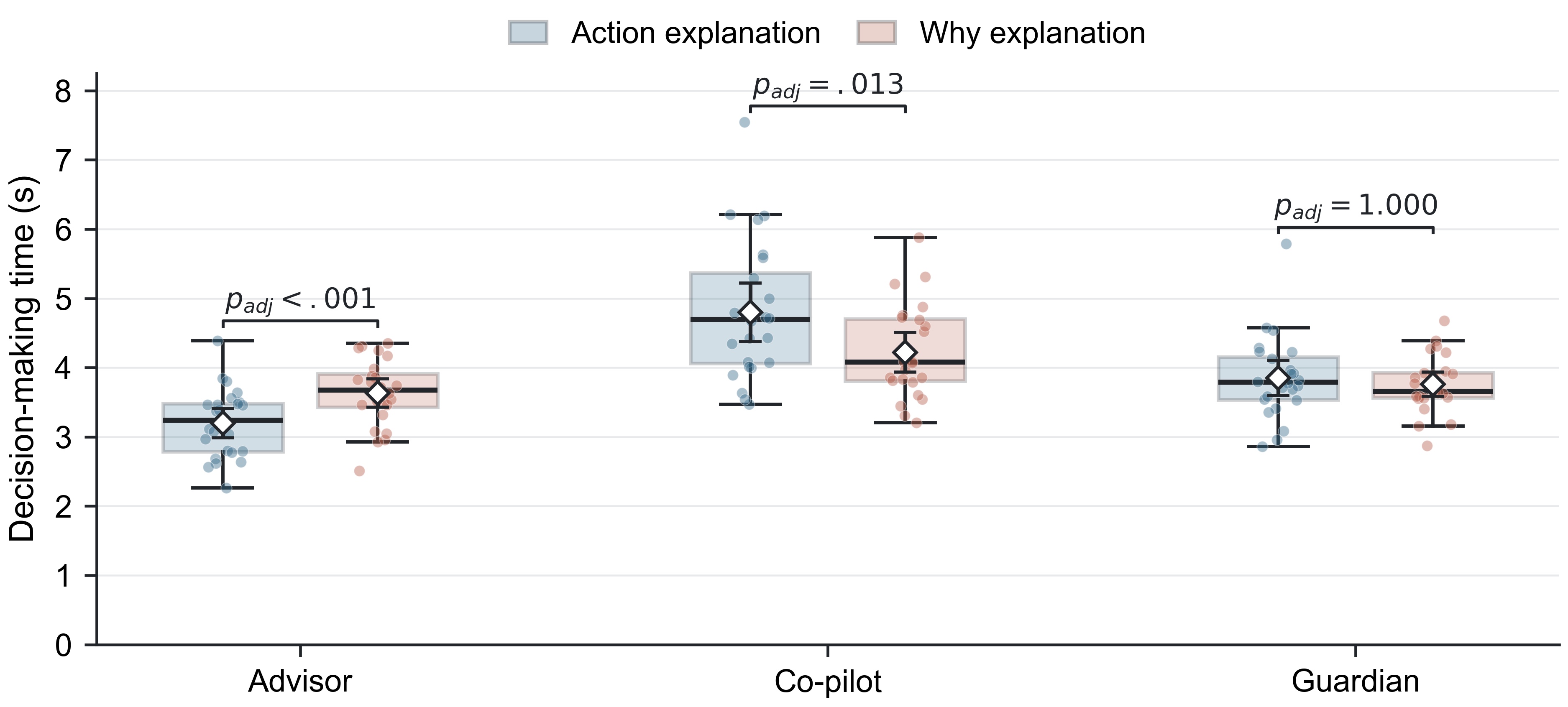}
\caption{Decision-making time as a function of AI role and explanation type.}
\label{fig:8}
\end{figure*}

There was also a significant AI role $\times$ explanation type interaction on decision-making time ($F$ (1.728, 39.753) = 15.654, $p$ \textless{} .001, $\eta_p^2$ = .405) (Fig.~\ref{fig:8}). Within Advisor, the ``action'' explanation ($M$ = 3.196 s, SEM = .102) produced shorter decision-making time than the ``why'' explanation ($M$ = 3.630 s, SEM = .099). Within Co-pilot, the ``why'' explanation ($M$ = 4.220 s, SEM = .138) produced shorter decision-making time than the ``action'' explanation ($M$ = 4.799 s, SEM = .205). Within Guardian, no difference was found between the ``why'' explanation ($M$ = 3.759 s, SEM = .085) and the ``action'' explanation ($M$ = 3.851 s, SEM = .124). Bonferroni-adjusted simple-effect tests confirmed that the ``why'' explanation was slower than the ``action'' explanation within Advisor (MD = .434 s, 95\% CI [.263, .605], $t$ (23) = 6.562, adjusted $p$ \textless{} .001, $d_z$ = 1.339) but faster within Co-pilot (MD = -.579 s, 95\% CI [-1.050, -.109], $t$ (23) = -3.178, adjusted $p$ = .013, $d_z$ = -.649). The explanation types did not differ within Guardian (MD = -.091 s, 95\% CI [-.427, .244], $t$ (23) = -.704, adjusted $p$ = 1.000, $d_z$ = -.144).

\subsubsection{Baseline comparison}\label{sec:3.1.2}

The repeated-measures ANOVA comparing the three AI roles with the sound-only Baseline showed a significant effect of condition on average decision-making time ($F$ (2.372, 54.556) = 30.586, $p$ \textless{} .001, $\eta_p^2$ = .571; Fig.~\ref{fig:7}(a)). Advisor ($M$ = 3.413 s, SEM = .095) was faster than Co-pilot ($M$ = 4.510 s, SEM = .149; MD = -1.097 s, 95\% CI [-1.442, -.751], $t$ (23) = -9.160, adjusted $p$ \textless{} .001, $d_z$ = -1.870), Guardian ($M$ = 3.805 s, SEM = .084; MD = -.392 s, 95\% CI [-.664, -.120], $t$ (23) = -4.153, adjusted $p$ = .002, $d_z$ = -.848), and Baseline ($M$ = 4.532 s, SEM = .166; MD = -1.119 s, 95\% CI [-1.518, -.719], $t$ (23) = -8.085, adjusted $p$ \textless{} .001, $d_z$ = -1.650). Guardian was faster than Co-pilot (MD = -.705 s, 95\% CI [-1.152, -.257], $t$ (23) = -4.544, adjusted $p$ \textless{} .001, $d_z$ = -.927) and Baseline (MD = -.727 s, 95\% CI [-1.221, -.233], $t$ (23) = -4.249, adjusted $p$ = .002, $d_z$ = -.867). Co-pilot and Baseline did not differ (MD = -.022 s, 95\% CI [-.461, .416], $t$ (23) = -.147, adjusted $p$ = 1.000, $d_z$ = -.030).

\subsection{Decision-making accuracy}\label{sec:3.2}

\subsubsection{Full-factorial analysis}\label{sec:3.2.1}

The factorial binomial GLMM showed a significant main effect of AI role on decision-making accuracy ($\chi^2$ (2) = 29.699, $p$ \textless{} .001). There was no significant main effect of explanation type ($\chi^2$ (1) = 1.349, $p$ = .245) or directional semantics ($\chi^2$ (1) = .984, $p$ = .321). There was no significant interaction effect of AI role $\times$ explanation type ($\chi^2$ (2) = 1.613, $p$ = .446), AI role $\times$ directional semantics ($\chi^2$ (2) = .850, $p$ = .654), explanation type $\times$ directional semantics ($\chi^2$ (1) = 2.051, $p$ = .152), or the three-way interaction ($\chi^2$ (2) = .776, $p$ = .678).

Decision-making accuracy approached the ceiling in the Advisor and Guardian conditions, and all 48 responses in the Guardian-why-contralateral cell were correct. This ceiling made the ordinary saturated maximum-likelihood model nearly unidentifiable and reduced sensitivity for detecting effects involving explanation type and directional semantics.

\subsubsection{Baseline comparison}\label{sec:3.2.2}

\begin{figure*}[!t]
\centering
\includegraphics[width=0.82\linewidth]{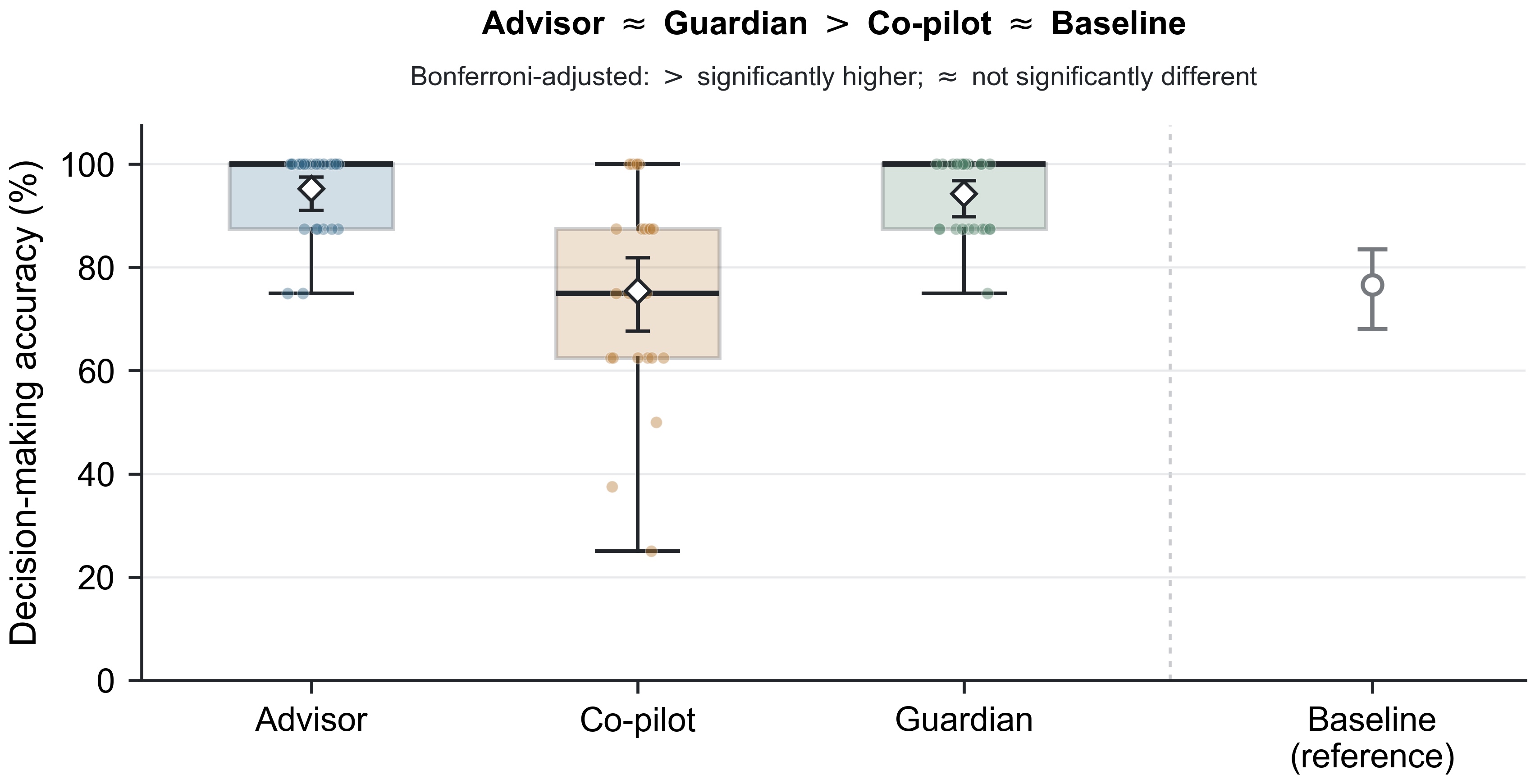}
\caption{Decision-making accuracy across the three AI roles and Baseline.}
\label{fig:9}
\end{figure*}

The GLMM comparing the three AI roles with the sound-only Baseline showed a significant effect of condition ($\chi^2$ (3) = 46.569, $p$ \textless{} .001; Fig.~\ref{fig:9}). Model-based predicted probabilities were .952 (95\% CI [.911, .975]) for Advisor, .754 (95\% CI [.677, .818]) for Co-pilot, .943 (95\% CI [.898, .968]) for Guardian, and .767 (95\% CI [.680, .835]) for Baseline. Bonferroni-adjusted comparisons showed that Advisor had higher odds of a correct response than Co-pilot (OR = 6.500, 95\% CI [2.473, 17.084], adjusted $p$ \textless{} .001) and Baseline (OR = 6.076, 95\% CI [2.227, 16.575], adjusted $p$ \textless{} .001). Guardian also had higher odds than Co-pilot (OR = 5.346, 95\% CI [2.162, 13.218], adjusted $p$ \textless{} .001) and Baseline (OR = 4.997, 95\% CI [1.943, 12.855], adjusted $p$ \textless{} .001). Advisor did not differ from Guardian (OR = 1.216, 95\% CI [.380, 3.886], adjusted $p$ = 1.000), and Co-pilot did not differ from Baseline (OR = .935, 95\% CI [.472, 1.852], adjusted $p$ = 1.000).

\subsection{Trust}\label{sec:3.3}

There was a significant main effect of AI role ($F$ (1.882, 43.276) = 3.966, $p$ = .028, $\eta_p^2$ = .147) on trust. Across AI roles, Advisor ($M$ = 5.535, SEM = .169) received higher trust ratings than Co-pilot ($M$ = 5.207, SEM = .192), while Guardian had an intermediate rating ($M$ = 5.504, SEM = .164). Bonferroni-adjusted comparisons showed that Advisor received higher trust ratings than Co-pilot (MD = .328, 95\% CI [.027, .629], $t$ (23) = 2.813, adjusted $p$ = .030, $d_z$ = .574). Advisor did not differ from Guardian (MD = .031, 95\% CI [-.290, .352], $t$ (23) = .251, adjusted $p$ = 1.000, $d_z$ = .051), and Co-pilot did not differ from Guardian (MD = -.297, 95\% CI [-.667, .073], $t$ (23) = -2.071, adjusted $p$ = .149, $d_z$ = -.423). There was no significant main effect of explanation type ($F$ (1, 23) = .029, $p$ = .866, $\eta_p^2$ = .001) or directional semantics ($F$ (1, 23) = 2.624, $p$ = .119, $\eta_p^2$ = .102) on trust.

\subsection{Anxiety}\label{sec:3.4}

There was a significant main effect of AI role ($F$ (1.781, 40.958) = 4.985, $p$ = .014, $\eta_p^2$ = .178) on anxiety. Participants reported mean anxiety ratings of 2.987 (SEM = .189) under Co-pilot, 2.719 (SEM = .184) under Advisor, and 2.595 (SEM = .204) under Guardian. Bonferroni-adjusted comparisons showed that Co-pilot produced higher anxiety than Guardian (MD = .392, 95\% CI [.014, .770], $t$ (23) = 2.675, adjusted $p$ = .041, $d_z$ = .546). Advisor did not differ from Co-pilot (MD = -.268, 95\% CI [-.550, .013], $t$ (23) = -2.462, adjusted $p$ = .065, $d_z$ = -.503) or Guardian (MD = .124, 95\% CI [-.192, .440], $t$ (23) = 1.011, adjusted $p$ = .968, $d_z$ = .206). There was no significant main effect of explanation type ($F$ (1, 23) = .044, $p$ = .836, $\eta_p^2$ = .002) or directional semantics ($F$ (1, 23) = .368, $p$ = .550, $\eta_p^2$ = .016) on anxiety. There was a significant AI role $\times$ directional semantics interaction effect on anxiety ($F$ (1.855, 42.667) = 4.955, $p$ = .013, $\eta_p^2$ = .177). Within Co-pilot, ipsilateral semantics produced lower anxiety than contralateral semantics (MD = -.307, 95\% CI [-.576, -.039], $t$ (23) = -2.953, adjusted $p$ = .021, $d_z$ = -.603). The corresponding directional effect was not significant within Advisor or Guardian.

\subsection{Usefulness}\label{sec:3.5}

There was a significant main effect of AI role ($F$ (1.975, 45.418) = 4.280, $p$ = .020, $\eta_p^2$ = .157) on usefulness. Across AI roles, Advisor ($M$ = 4.233, SEM = .094) and Guardian ($M$ = 4.217, SEM = .114) received higher mean usefulness ratings than Co-pilot ($M$ = 3.997, SEM = .139). Bonferroni-adjusted comparisons showed that Advisor received higher usefulness ratings than Co-pilot (MD = .236, 95\% CI [.013, .460], $t$ (23) = 2.736, adjusted $p$ = .035, $d_z$ = .559). Advisor did not differ from Guardian (MD = .017, 95\% CI [-.229, .262], $t$ (23) = .175, adjusted $p$ = 1.000, $d_z$ = .036), and Co-pilot did not differ from Guardian (MD = -.220, 95\% CI [-.450, .010], $t$ (23) = -2.468, adjusted $p$ = .064, $d_z$ = -.504). There was no significant main effect of explanation type ($F$ (1, 23) = .027, $p$ = .871, $\eta_p^2$ = .001) or directional semantics ($F$ (1, 23) = .220, $p$ = .643, $\eta_p^2$ = .009) on usefulness.

\subsection{Satisfaction}\label{sec:3.6}

There was a significant main effect of AI role ($F$ (1.889, 43.438) = 5.355, $p$ = .009, $\eta_p^2$ = .189) on satisfaction. Across AI roles, Advisor ($M$ = 3.919, SEM = .142) and Guardian ($M$ = 3.934, SEM = .151) received higher mean satisfaction ratings than Co-pilot ($M$ = 3.655, SEM = .167). Bonferroni-adjusted comparisons showed that Advisor received higher satisfaction ratings than Co-pilot (MD = .264, 95\% CI [.004, .524], $t$ (23) = 2.625, adjusted $p$ = .045, $d_z$ = .536), and Guardian received higher satisfaction ratings than Co-pilot (MD = .279, 95\% CI [.014, .543], $t$ (23) = 2.722, adjusted $p$ = .036, $d_z$ = .556). Advisor did not differ from Guardian (MD = -.014, 95\% CI [-.230, .201], $t$ (23) = -.172, adjusted $p$ = 1.000, $d_z$ = -.035). There was no significant main effect of explanation type ($F$ (1, 23) = .015, $p$ = .904, $\eta_p^2$ = .001) or directional semantics ($F$ (1, 23) = .018, $p$ = .893, $\eta_p^2$ = .001) on satisfaction.

\subsection{Workload}\label{sec:3.7}

There was a significant main effect of AI role ($F$ (1.625, 37.374) = 17.025, $p$ \textless{} .001, $\eta_p^2$ = .425) on perceived workload. Participants reported the lowest workload under Guardian ($M$ = 12.724, SEM = 1.715), followed by Advisor ($M$ = 16.932, SEM = 2.128), with the highest workload under Co-pilot ($M$ = 20.672, SEM = 2.322). Bonferroni-adjusted comparisons showed that Advisor produced lower workload than Co-pilot (MD = -3.740, 95\% CI [-6.808, -.671], $t$ (23) = -3.147, adjusted $p$ = .014, $d_z$ = -.642) but higher workload than Guardian (MD = 4.208, 95\% CI [1.142, 7.275], $t$ (23) = 3.543, adjusted $p$ = .005, $d_z$ = .723). Co-pilot also produced higher workload than Guardian (MD = 7.948, 95\% CI [3.666, 12.229], $t$ (23) = 4.793, adjusted $p$ \textless{} .001, $d_z$ = .978). There was no significant main effect of explanation type ($F$ (1, 23) = .237, $p$ = .631, $\eta_p^2$ = .010) or directional semantics ($F$ (1, 23) = .002, $p$ = .964, $\eta_p^2$ = .000) on workload.

\subsection{Temporal-order sensitivity analysis}\label{sec:3.8}

The joint linear and quadratic trial-position effects were significant after Holm correction for anxiety ($\chi^2$ (2) = 15.64, adjusted $p$ = .002) and workload ($\chi^2$ (2) = 30.83, adjusted $p$ \textless{} .001). Both outcomes showed positive quadratic effects, indicating higher ratings near the beginning and end of the session than during the middle. There was no corrected temporal-order effect for decision-making time, decision-making accuracy, trust, usefulness, or satisfaction (Table~\ref{tab:5}). Adjusting for chronological trial position did not change the statistical significance of any factorial effect for the continuous outcomes or decision-making accuracy.

\begin{table}[!t]
\centering
\caption{Joint temporal-order sensitivity tests.}
\label{tab:5}
\begingroup
\footnotesize
\setlength{\tabcolsep}{3pt}
\renewcommand{\arraystretch}{1.15}
\begin{tabular}{@{}>{\raggedright\arraybackslash}p{0.51\linewidth}>{\raggedright\arraybackslash}p{0.24\linewidth}>{\raggedright\arraybackslash}p{0.18\linewidth}@{}}
\toprule
\textbf{Outcome} & \textbf{Joint temporal $\chi^2$ (2)} & \textbf{Holm-adjusted p} \\
\midrule
Decision-making time & 5.11 & .233 \\
Decision-making accuracy & 2.39 & .303 \\
Trust & 6.15 & .185 \\
Anxiety & 15.64 & .002 \\
Usefulness & 6.61 & .183 \\
Satisfaction & 4.67 & .233 \\
Workload & 30.83 & \textless{} .001 \\
\bottomrule
\end{tabular}
\par\smallskip\begin{minipage}{\linewidth}\footnotesize
Note. The joint test included standardized linear and quadratic chronological trial-position terms. Continuous outcomes used likelihood-ratio tests, and decision-making accuracy used a joint Wald test. Holm corrections were applied across the seven outcomes.
\end{minipage}
\endgroup
\end{table}

\section{Discussion}\label{sec:4}

This study compared three AI-role authority-ability configurations (Advisor, Co-pilot, and Guardian) and examined how explanation type and spoken directional semantics affected decision-making time, decision-making accuracy, and subjective ratings during SAE Level 3 takeovers. Advisor was associated with the shortest decision-making time, Guardian with the next shortest time, and Co-pilot with the longest time. Advisor and Guardian were also associated with higher decision-making accuracy and satisfaction than Co-pilot. Guardian was associated with the lowest workload, whereas Co-pilot was associated with the highest. Explanation type had no overall main effect, but its effect on decision-making time depended on AI role. Action explanations were faster under Advisor, why explanations were faster under Co-pilot, and the two explanation types did not differ under Guardian. Ipsilateral semantics were associated with shorter decision-making time than contralateral semantics. The effect of directional semantics on anxiety also depended on AI role, with ipsilateral semantics producing lower anxiety than contralateral semantics only under Co-pilot. The temporal-order sensitivity analysis showed that, although anxiety and workload varied across trial position, adjusting for temporal order did not change the significance of any factorial effect.

\subsection{AI role}\label{sec:4.1}

The findings of this study on AI-role configurations indicated that Advisor AI had the shortest decision-making time, while Co-pilot AI had the longest. Advisor AI and Guardian AI also had higher decision-making accuracy than Co-pilot AI in the takeover context. A possible explanation is that Advisor AI (e.g., ``Hazard ahead, left/right lane clear,'' or ``Avoid the left/right immediately.''), which provides informational warnings that directly alert users, either suggests the required lane change action or indicates current traffic conditions. This message format showed which lane was open and directly corresponded to the lane choice, which may have made it easier for the driver to choose the correct action and may have further improved decision-making quality (e.g., reducing decision-making time and increasing decision-making accuracy) \cite{r166,r220}. Conversely, Co-pilot AI (e.g., ``Realign, left/right steering insufficient,'' or ``Hold steady, reducing left/right steering.'') provided information to correct the user's control input. These messages contained lower-level steering language (i.e., ``left/right steering insufficient'' or ``reducing left/right steering''), which differed from the typical lane-choice goal that drivers receive from a system (i.e., ``Change to the left/right lane immediately''). Therefore, drivers may have needed to map the lower-level steering information onto the lane-choice response, which may help explain the longer decision-making time and lower accuracy \cite{r172,r191}. After receiving the message from Guardian AI, the driver still needed to verify lane availability, but the authority assigned to Guardian enabled it to execute the vehicle maneuver, which may have reduced the control-transition and coordination demands required to make an accurate decision \cite{r219}. Although Guardian AI did not require the driver to perform the maneuver manually, both decision-making measures were based on the indicator-button response, which occurred before either the automated or manual lane change. Because the indicator response preceded the role-dependent automation-disengagement and maneuver-execution actions, decision-making time ended at the same lane-choice response in all three configurations. Even so, the measure reflects the complete configurations as implemented and does not isolate message interpretation from the other features of each authority-ability configuration. For Guardian, the absence of immediate driving-related input from the driver during the critical scenario may have reduced perceived urgency and contributed to its longer decision-making time than Advisor AI \cite{r262}. In addition, compared with Baseline, which did not include an integrated AI-driven system, Co-pilot was the only AI agent that did not show better performance on either decision-making measure.

The AI role $\times$ explanation type interaction further suggests that the effect of explanation content depended on how it complemented the information provided by each AI-role configuration. Decision-making time was shorter with an ``action'' explanation than with a ``why'' explanation within Advisor, whereas Co-pilot showed the opposite pattern. In other words, the information provided by Advisor AI may already have contained a ``why'' component, and adding an ``action'' explanation provided a direct mapping from information to lane choice. Consistent with this, the ``why'' message could have partially overlapped with the information provided by Advisor AI, resulting in a lower impact \cite{r195}. Similarly, the ``why'' explanations within Co-pilot AI provided context about why the correction of the user's input was needed, which may have reduced ambiguity and shortened decision-making time relative to the ``action'' explanation \cite{r192}.

Advisor AI received higher trust and usefulness ratings than Co-pilot AI, and Guardian AI was associated with lower anxiety than Co-pilot AI. A possible explanation is that the shared-control arrangement implemented by Co-pilot AI may have created responsibility ambiguity, which in turn may have led to conflict during the transfer of control \cite{r163,r167}. This conflict may have reduced drivers' ease of decision-making and may also help explain why we observed higher workload and anxiety under this AI role. Conversely, as mentioned earlier, the assigned authority allowed Guardian AI to execute the maneuver, which may have reduced the potential coordination burden that might otherwise have increased driver workload \cite{r237}. Similarly, Advisor AI may have reduced such conflict by only providing informational messages, leaving the driver to perform the action independently. This division of responsibility may have reduced uncertainty and lowered coordination costs, which may help explain the higher trust and usefulness ratings \cite{r196,r247}.

The objective and subjective findings further suggest that the timing of AI communication could also be important. Advisor AI was similar to current warning systems that provide warning information when the system detects danger, which drivers may already be familiar with \cite{r130,r203,r261}. Given the capabilities of Co-pilot AI, its benefit might be more noticeable when drivers are already engaged in the manual driving task. Therefore, the less favorable results observed under Co-pilot AI may have been caused by a mismatch between event timing and AI capabilities. Future work could examine the effects of these three AI-role configurations across takeover stages (e.g., before takeover, during takeover, after takeover) and integrate capabilities from different AI-role configurations into the phases where they perform best \cite{r154,r170,r209}.

\subsection{Explanation type}\label{sec:4.2}

Explanation type had no overall main effect on any outcome. Since explanation content can strongly shape how drivers understand system behavior, it remains important to determine whether these results hold in more realistic, non-binary takeover situations. In real traffic scenarios, more complex conditions, such as uncertain gaps, merge pressure, speed constraints, and conflicting route goals between the driver and the system, might occur \cite{r183,r184}. In this context, the ``why'' explanation might provide more cues to drivers, highlighting potential risks and their reasons. In contrast, the ``action'' explanation might offer an action plan for what to do now and in what sequence \cite{r194}. Presenting the same situation with different information priorities might engage different cognitive routes and influence driving choices and the safety of the outcomes. Additionally, the current study examined a critical scenario in which drivers had to make a decision under time pressure. It is important to determine whether this pattern continues in the post-takeover scenario, where the situation is stable, and drivers do not need to perform driving tasks urgently \cite{r170,r199,r260}. For example, during the post-takeover stage, immediately after a critical event, a companion-style or emotionally supportive explanation, such as ``There was a vehicle that pulled out a minute ago. You are safe now. Do you want to pull over and rest?'' might be needed. Capturing drivers' attitudes at different stages helps identify when to present these explanation types. This can provide more complete insights into how explanations can be improved or adjusted to better meet drivers' needs across the entire event, including helping to build a clear mental model and supporting continued use.

\subsection{Directional semantics}\label{sec:4.3}

Ipsilateral semantics were associated with shorter decision-making time than contralateral semantics. These findings align with SRC, under which performance improves when the stimulus and the required response share the same left or right semantics and declines when they are opposite in direction \cite{r223,r224}. In the current study, the directional information conveyed through ipsilateral semantics matched the required action direction (e.g., ``Change to the left/right lane immediately''). This direct mapping may have supported the interpretation process and reduced the time needed to decide \cite{r202,r203,r206}. Conversely, contralateral semantics, which provided opposite directional information, may have required the driver to process the reversed message before performing the action, which may explain the slower responses.

The result for decision-making time conflicts with some earlier studies that reported no differences between the two types of semantics \cite{r181,r220}. A possible explanation is that our study used clear verbal instructions instead of spatial cues (e.g., vibrating or playing a sound on a specific side), which directly informed drivers how to choose a lane. Although spatial cues in previous studies were found to help redirect attention, they might also create uncertainty in conveying the message, such as ``which side had the vibration?'' or ``which side did the sound come from?'' \cite{r150,r164}. Future studies should assess whether speech-based directional semantics help drivers detect and prioritize critical elements and build and maintain situation awareness in complex traffic conditions with a high information load, which would support effective use in real-world traffic, where timely understanding guides safer choices \cite{r241}.

\subsection{Limitations}\label{sec:4.4}

This study has several limitations. First, each AI role was implemented as a complete authority-ability configuration in which control authority, required driver actions, and role-specific message wording varied together. For example, Guardian did not require drivers to disengage automation or perform the maneuver manually, while Advisor and Co-pilot required both actions. The why and action messages were also phrased to reflect each role's function. Therefore, the design cannot isolate the contribution of any single component. All reported AI-role effects reflect differences among the three configurations as implemented and cannot be attributed specifically to control authority, task demand, or message wording alone. Differences in the required actions are particularly relevant to workload and decision-making time, so these results should be interpreted with caution. The explanation-type results require the same caution because message wording varied across roles. Second, accuracy approached the ceiling under Advisor and Guardian, reducing sensitivity to effects involving explanation type and directional semantics. Future studies could increase decision difficulty through ambiguous or late-resolving lane availability, additional surrounding vehicles, or shorter lead times. They could also use graded measures of decision quality, such as lane-choice confidence or corrective inputs after the initial choice, to retain variance when binary accuracy reaches the ceiling. Process measures such as eye tracking could provide additional evidence about the decision process, which binary accuracy cannot capture.

Third, the current study used a single, simple scenario under basic traffic conditions as the testing environment. To support the wide adoption of AI-role configurations, these configurations should be evaluated and verified comprehensively across more diverse and complex scenarios. Therefore, future studies should consider increasing scenario complexity by adding realistic conditions such as traffic density, weather, and road type. Finally, as part of this preliminary investigation of the effects of the triadic Human-AI collaboration framework, this study adopted the widely used auditory HMI as the communication system. However, the impact of other unimodal (e.g., visual or tactile) and multimodal HMIs, which have shown effectiveness in conveying system information in previous studies, remains unclear. Future research should consider incorporating these HMIs to further identify the optimal communication strategy.

\section{Conclusion}\label{sec:5}

This study examined the effects of three AI-role authority-ability configurations, explanation type, and directional semantics on objective takeover performance and subjective ratings during SAE Level 3 takeovers. Advisor was associated with the shortest decision-making time, followed by Guardian, whereas Co-pilot was associated with the longest. In comparisons with Baseline, Advisor and Guardian produced shorter decision-making time and higher decision-making accuracy, whereas Co-pilot did not differ from Baseline on either outcome. Advisor and Guardian received higher satisfaction ratings than Co-pilot; Advisor received higher trust and usefulness ratings than Co-pilot; and Guardian was associated with lower anxiety than Co-pilot. Guardian was associated with the lowest workload, whereas Co-pilot was associated with the highest. The effect of explanation type depended on AI role. Action explanations shortened decision-making time within Advisor, why explanations shortened decision-making time within Co-pilot, and the two explanation types did not differ within Guardian. Ipsilateral semantics were associated with shorter decision-making time than contralateral semantics. The findings from this study offer guidance on defining how technical AI abilities are exercised under different degrees of control authority in automated driving, which could enhance collaboration between humans and AI. This work also provides guidance for developing communication strategies that clarify system capabilities for users, which may further enhance the adoption of AI-driven systems. In practice, an Advisor configuration that delivers takeover support through a combination of an action explanation and ipsilateral semantics (e.g., ``Change to the left/right lane immediately.'') may be a promising option for general takeover scenarios. This pairing combines design elements associated with shorter decision-making time. In contexts where the vehicle is permitted to execute the maneuver, a Guardian configuration with ipsilateral semantics (e.g., "Engaging, steering left/right for safety.") is a defensible alternative. Guardian did not differ significantly from Advisor in decision-making accuracy and had higher decision-making accuracy and shorter decision-making time than Co-pilot and Baseline. Guardian was also associated with lower anxiety than Co-pilot. This recommendation is based on the decision-making and anxiety results. The lower workload observed under Guardian is not treated as supporting evidence because it partly reflects the reduced manual demand inherent in that configuration (Section~\ref{sec:4.4}). Across both configurations, support for ipsilateral semantics comes from the results for decision-making time. The anxiety benefit of ipsilateral semantics appeared only under Co-pilot. These insights might help guide the integration of AI in the transportation context, informing the development of next-generation AI-driven systems.

\FloatBarrier

\end{document}